\documentclass[aps,prb,reprint,superscriptaddress,floatfix,longbibliography]{revtex4-2}

\usepackage{latexsym}
\usepackage{graphicx}
\usepackage{lmodern}

\usepackage{amsmath}
\usepackage{amssymb}
\usepackage{bm}
\usepackage{braket}
\usepackage{mathtools}
\usepackage{hyperref}
\usepackage{url}

\usepackage{booktabs}
\usepackage{capt-of}
\usepackage{enumitem}
\usepackage{multirow}

\usepackage[caption=false]{subfig}

\usepackage{indentfirst}

\hypersetup{
  colorlinks=true,
  citecolor=blue,
  linkcolor=magenta,
  urlcolor=magenta,
}

\newcommand{\AffHandai}{Graduate School of Engineering Science, The University of Osaka, 1-3 Machikaneyama, Toyonaka, Osaka 560-8531, Japan}
\newcommand{\AffQIQB}{Center for Quantum Information and Quantum Biology, The University of Osaka, 1-2 Machikaneyama, Toyonaka, Osaka 560-0043, Japan}
\newcommand{\AffQunasys}{QunaSys Inc., Aqua Hakusan Building 9F, 1-13-7 Hakusan, Bunkyo, Tokyo 113-0001, Japan}

\begin{document}

\title{Evaluating higher-order product formulae for \\ molecular ground-state energy estimation}

\author{Hiromu Abe}
\affiliation{\AffHandai}

\author{Keita Kanno}
\affiliation{\AffQunasys}

\author{Ryosuke Kimura}
\affiliation{\AffHandai}

\author{Masahiko Kamoshita}
\affiliation{\AffHandai}

\author{Kosuke Mitarai}
\affiliation{\AffHandai}
\affiliation{\AffQIQB}

\begin{abstract}
We evaluate deterministic higher-order product formulae for molecular ground-state energy estimation. Motivated by recent fault-tolerant architectures in which non-Clifford operations may be generated more locally and cheaply than in conventional assumptions, we re-examine such formulae as practical candidates for quantum chemistry. Using one-dimensional hydrogen chains from $\mathrm{H}_2$ to $\mathrm{H}_{15}$ as benchmarks, we estimate both the total gate count and the depth of $R_Z$-rotation layers required to reach a target energy error. To make this comparison feasible at larger system sizes, we use a perturbative method to estimate the eigenvalue error induced by each product formula and thereby evaluate the cost of the corresponding phase-estimation procedure. Among the previously considered formulae, the eighth-order construction introduced by Morales \textit{et al.} [M. E. S. Morales \textit{et al.}, ``Greatly improved higher-order product formulae for quantum simulation,'' arXiv:2210.15817v2 (2024)] minimizes both cost metrics in the benchmark at a chemically relevant target error. We also find that increasing the formal order does not automatically reduce the total cost: near chemical accuracy, the tenth-order formula introduced in the same work can be less efficient than the eighth-order one. Motivated by this observation, we construct a new fourth-order formula; it achieves the lowest total gate count among the formulae considered for all H-chain instances near chemical accuracy and over much of the \(0.1\)--\(10\,\mathrm{mHa}\) target-error window for most instances, while also reducing the $R_Z$-layer depth. These results clarify how deterministic higher-order product formulae should be selected for molecular ground-state energy estimation.

\end{abstract}

\maketitle

\section{Introduction}
Quantum simulation of quantum mechanical systems is an application of quantum computers. It has the potential to solve problems that are intractable on classical computers, ranging from quantum chemistry to condensed matter and high-energy physics. Among the various tasks in quantum simulation, obtaining the eigenvalues of the system Hamiltonian is important, as they correspond to the energy levels of the system. Classical algorithms typically require computational resources that grow exponentially with system size. In contrast, quantum algorithms can find these eigenvalues in polynomial time \cite{Kitaev1995PhaseEstimation,AbramsLloyd1999Eigenvalues}.

To this end, a variety of quantum algorithms have been developed. Product formulae, also known as Trotter--Suzuki decompositions, have served as methods for simulating Hamiltonian time evolution \cite{Trotter1959,Suzuki1976}. More recently, methods based on block encoding and quantum singular value transformation (QSVT) have attracted attention \cite{LowChuang2019,Gilyen2019} since they can exponentially improve the cost of implementing a fixed-time Hamiltonian simulation step with respect to the error compared with product formulae. They are therefore considered suitable for high-precision energy estimation.
However, the practical advantage of such algorithms has relied on the assumption that non-Clifford operations, such as T gates, are costly due to the resource overhead of magic-state distillation \cite{BravyiKitaev2005MSD}.

This paradigm is beginning to shift with recent advances in fault-tolerant quantum computing architectures. Proposals such as magic-state cultivation \cite{Gidney2024Cultivation}, zero-level distillation \cite{Hirano_2024ZLD}, and the STAR architecture \cite{Akahoshi2024STAR} suggest the possibility of generating magic states locally with fewer resources than before. In such settings, algorithms that exploit the locality of the Hamiltonian can become more resource-efficient than QSVT, which requires more global structures. These changes motivate a re-evaluation of the performance of product formulae for practical quantum simulation.

Recent work on product-formula-based Hamiltonian simulation has developed along several directions. In the deterministic setting, Morales \textit{et al.} \cite{Mauro} revisited Yoshida's classical constructions \cite{Yoshida} and numerically optimized the product-formula parameters. A subsequent study by the same authors further sharpened this line of work by adopting eigenvalue-based error metrics and reporting improved higher-order constructions \cite{MoralesQIC2025}. In parallel, randomized and partially randomized product-formula methods have emerged as another promising direction for phase-estimation-based quantum chemistry \cite{Gunther2025PartiallyRandomized,Kanasugi2026EarlyFTQC}.

\begin{figure}[!tb]
    \centering
    \includegraphics[width=\linewidth]{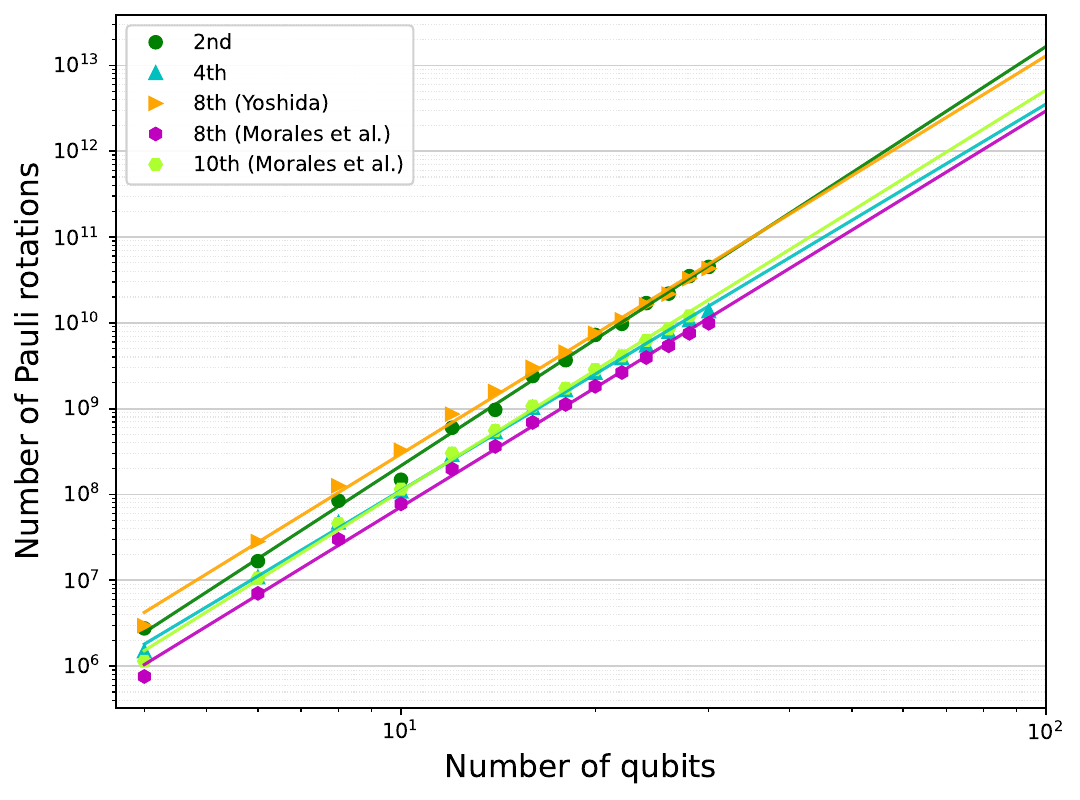}
    \caption{For H-chains at the target energy error $\varepsilon_{\mathrm{E}} = 1.6\times 10^{-4}\,\mathrm{Ha}$, we plot the total gate count $F$ (number of Pauli rotations) as dots for each qubit count. The colored straight lines show the fitted scaling on logarithmic scales for both axes. Legend entries are, in order, second order (green), conventional fourth order (cyan; labeled ``4th'' in the figure), eighth order by Yoshida (orange), eighth order by Morales \textit{et al.} (magenta), and tenth order by Morales \textit{et al.} (greenyellow).}
    \label{fig:extrapolation_total_gate_count}
\end{figure}

\begin{figure}[!tb]
    \centering
    \includegraphics[width=\linewidth]{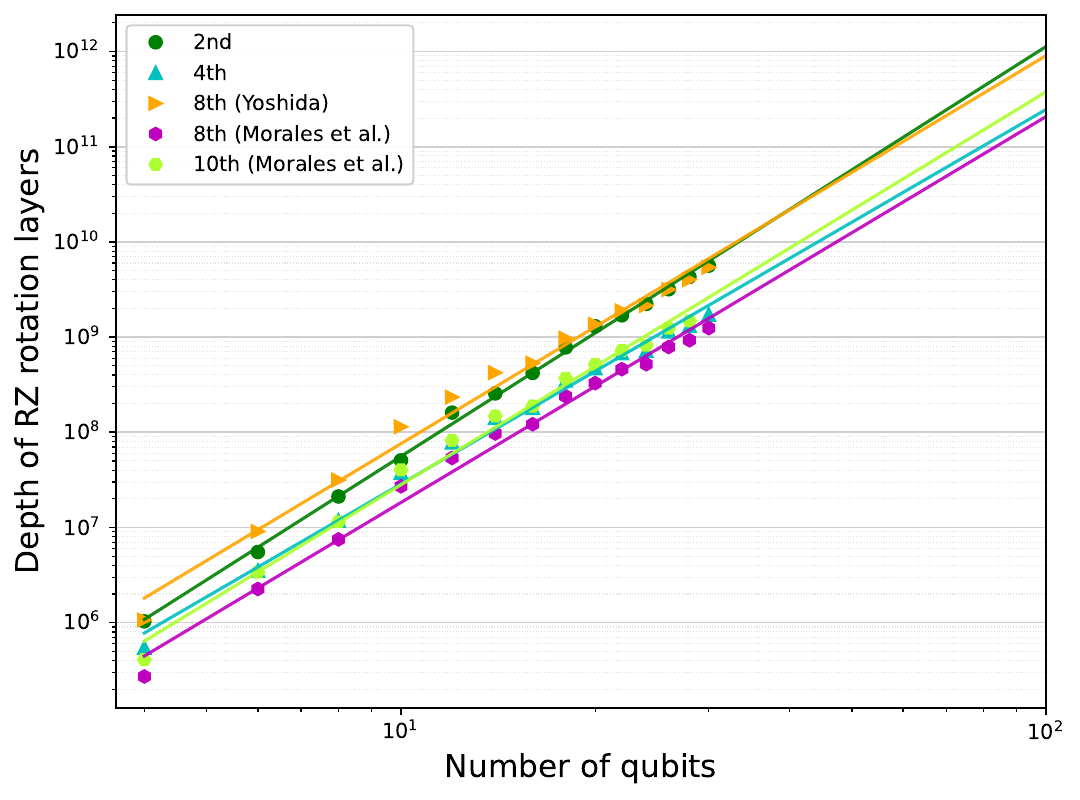}
    \caption{For H-chains at the target energy error $\varepsilon_{\mathrm{E}} = 1.6\times 10^{-4}\,\mathrm{Ha}$, we plot the depth of $R_Z$-rotation layers as dots for each qubit count. The colored straight lines show the fitted scaling on logarithmic scales for both axes. Legend entries are, in order, second order (green), conventional fourth order (cyan; labeled ``4th'' in the figure), eighth order by Yoshida (orange), eighth order by Morales \textit{et al.} (magenta), and tenth order by Morales \textit{et al.} (greenyellow).}
    \label{fig:extrapolation_rz_layer_depth}
\end{figure}

\begin{figure}[!tb]
    \centering
    \includegraphics[width=\linewidth]{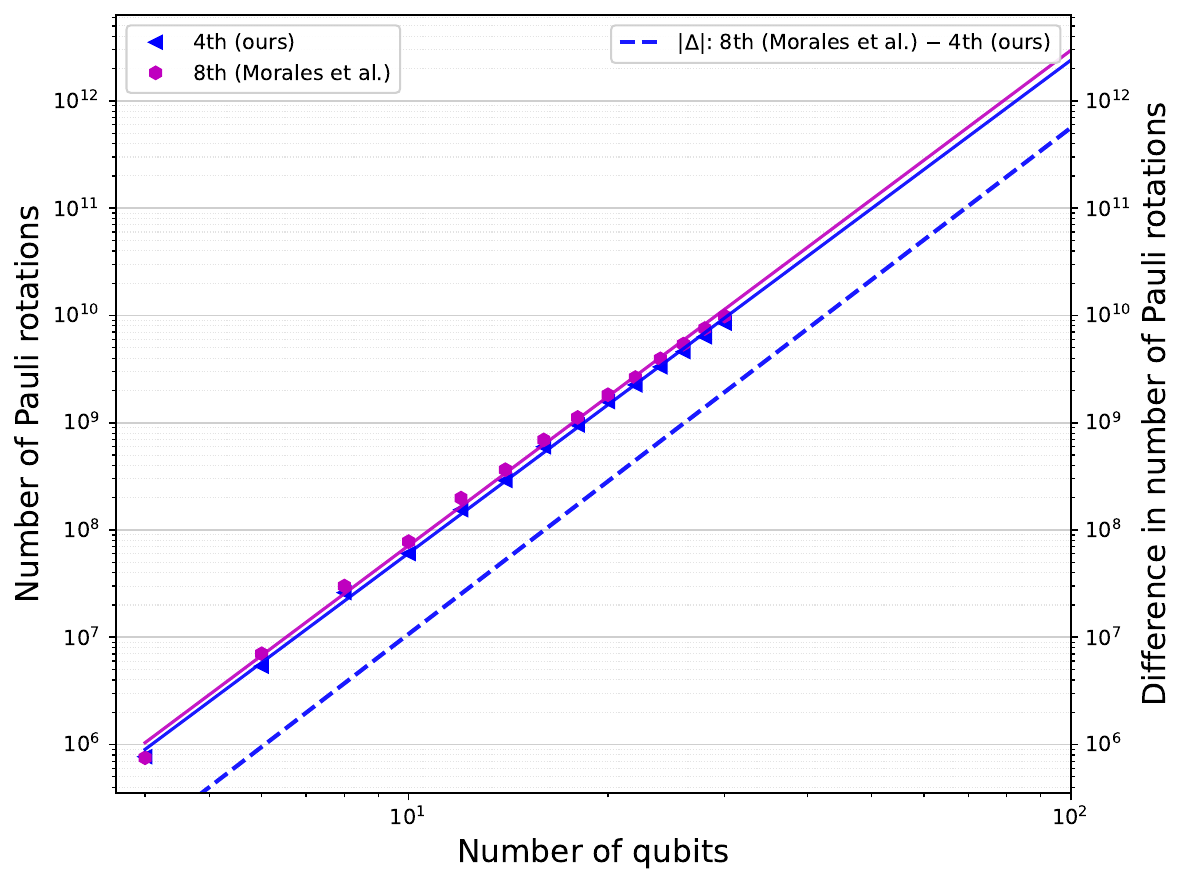}
    \caption{As in Fig.~\ref{fig:extrapolation_total_gate_count}, we plot the total gate count $F$ at the target energy error $\varepsilon_{\mathrm{E}} = 1.6\times 10^{-4}\,\mathrm{Ha}$ together with the fitted scaling on log--log axes. We compare the eighth-order product formula by Morales \textit{et al.} (magenta, ``8th (Morales \textit{et al.})'') with our newly constructed fourth-order product formula (blue, ``4th (ours)''). The dashed curve shows the absolute difference between the two estimated total gate counts, plotted on the right axis.}
    \label{fig:extrapolation_total_gate_count_fourth_vs_eighth}
\end{figure}

\begin{figure}[!tb]
    \centering
    \includegraphics[width=\linewidth]{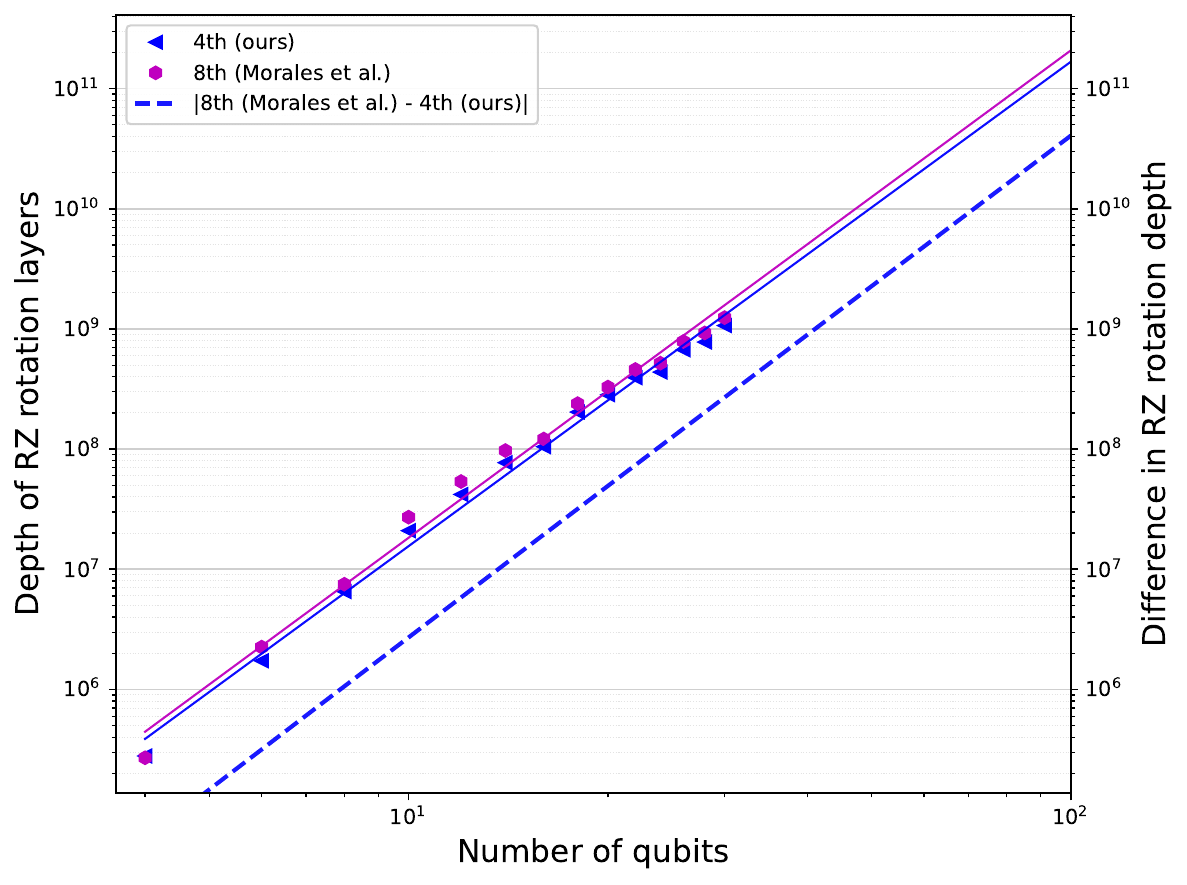}
    \caption{As in Fig.~\ref{fig:extrapolation_rz_layer_depth}, we plot the depth of $R_Z$-rotation layers at the target energy error $\varepsilon_{\mathrm{E}} = 1.6\times 10^{-4}\,\mathrm{Ha}$ together with the fitted scaling on log--log axes. We compare our new fourth-order product formula (blue, ``4th (ours)'') with the eighth-order product formula by Morales \textit{et al.} (magenta, ``8th (Morales \textit{et al.})''). The dashed curve shows the absolute difference between the two estimated depths of $R_Z$-rotation layers, plotted on the right axis.}
    \label{fig:extrapolation_rz_layer_depth_fourth_vs_eighth}
\end{figure}

In this work, we focus on the deterministic higher-order product formulae introduced by Morales \textit{et al.} \cite{Mauro} and quantitatively assess their suitability for molecular ground-state energy estimation. Although Ref.~\cite{MoralesQIC2025} has examined their performance for fermionic Hamiltonians, its numerical tests have been restricted to randomly generated all-to-all interacting six-orbital models. This setting does not directly reveal which formulae are preferable for molecular systems, nor how the associated resource costs scale with system size in practically relevant instances. We therefore analyze the errors induced by these product formulae and evaluate the resources required to compute ground-state energies for hydrogen chains. A perturbative strategy for estimating the eigenvalue error between the exact time-evolution operator and its product-formula approximation, which may be of independent interest, allows us to extend the analysis up to the 30-qubit scale. We then extrapolate the resulting trends to estimate the gate cost for larger systems. Although we do not consider randomized or partially randomized product-formula methods in the present work, recent analyses treat the deterministic Trotter error separately from the stochastic contributions in those settings \cite{Gunther2025PartiallyRandomized,Kanasugi2026EarlyFTQC}, so the same basic perspective should extend naturally to such methods.
Our analysis yields several findings. At a target precision of $0.16\,\mathrm{mHa}$, which is one order of magnitude tighter than chemical accuracy, the eighth-order product formula of Morales \textit{et al.} \cite{Mauro} is the best among the previously available deterministic formulae in both cost metrics: it minimizes the total number of Pauli rotations up to 100 qubits, and it also gives the smallest $R_Z$-layer depth in the extrapolation shown in Figs.~\ref{fig:extrapolation_total_gate_count} and~\ref{fig:extrapolation_rz_layer_depth}. Motivated by the observation that a higher formal order does not automatically lead to a lower overall cost, we further construct a new fourth-order formula. At the same target precision, this new formula gives the lowest total gate count for all H-chain instances considered; at the tighter $0.16\,\mathrm{mHa}$ target, it improves on the eighth-order formula for $\mathrm{H}_3$--$\mathrm{H}_{15}$ and in the extrapolated larger-system regime in both metrics, as shown in Figs.~\ref{fig:extrapolation_total_gate_count_fourth_vs_eighth} and~\ref{fig:extrapolation_rz_layer_depth_fourth_vs_eighth}.
These results provide a quantitative perspective on how deterministic higher-order product formulae should be selected for molecular ground-state energy estimation in practical fault-tolerant settings.

\section{Theoretical background}
First, in Sec.~\ref{sec:quantum_phase_estimation_and_product_formulae}, we review the background on quantum phase estimation and higher-order product formulae relevant to this work. Then, in Sec.~\ref{sec:total_cost_analysis}, we formulate the cost-analysis framework employed in our comparison.

\subsection{Quantum phase estimation and product formulae}\label{sec:quantum_phase_estimation_and_product_formulae}
\subsubsection{Quantum phase estimation}
Quantum phase estimation (QPE) \cite{Kitaev1995PhaseEstimation} is an algorithm to estimate eigenphases $\varphi$ of a unitary $U$ using controlled-version of $U$.
Throughout this work, we assume the idealized case in which the input state to QPE is the exact eigenstate of interest, namely the true molecular ground state. Thus, the resource estimates below do not include the cost of state preparation or the effect of imperfect ground-state overlap.
Let $\hat{\varphi}$ denote the phase estimated by QPE and $\varphi$ the true phase to be estimated. We define the estimation error of QPE as
\[
  \varepsilon = |\hat{\varphi} - \varphi|\, .
\]
The number of unitary applications required to estimate the phase with error $\varepsilon$, denoted $M(\varepsilon)$ is generally given by
\begin{equation}
    M(\varepsilon) = \frac{\beta}{\varepsilon},
    \label{eq:qpe_cost}
\end{equation}
for some constant $\beta$.
A theoretical upper-bound on $\beta$ is $\beta\leq 16\pi$ \cite{Reiher}.
In practice, however, a much smaller constant factor suffices to achieve error $\varepsilon$.
For resource analysis performed later, we use $\beta=1.56$, obtained from the numerical evaluation presented in Appendix~\ref{sec:beta_estimation}.

\subsubsection{Product formulae}
Suppose a Hamiltonian $H$ is expressed as a linear combination of $J$ terms $H_j$ with $j \in \{1,\ldots,J\}$:
\[
H=\sum^{J}_{j=1}{H_j}\, .
\]
Product formulae approximate the time-evolution operator $e^{iHt}$ via the product of $e^{iH_jt}$.
The second-order formula approximates the time evolution as
\begin{align}\label{eq:second_order_product_formula}
    S_2(t) &:= \prod^{J-1}_{p=1}{e^{\tfrac{1}{2}iH_{p}t}}e^{iH_{J}t}\prod^{J-1}_{q=1}e^{\tfrac{1}{2}iH_{J-q}t}\,.
\end{align}
It satisfies
\begin{align}\label{eq:second_order_product_formula_error}
    e^{iHt} = S_2(t) + \mathcal{O}(t^3)\, .
\end{align}
A common method to construct higher-order product formulae is Suzuki’s fractal decomposition \cite{Suzuki90fractal}. First, let $S_2(t)$ denote the second-order product formula in Eq.~\eqref{eq:second_order_product_formula}. A product formula of order $k=2\kappa$ with $\kappa \in \mathbb{N}$ can be constructed recursively as
\begin{equation}
    S_{2\kappa}(t)=S_{2\kappa-2}(s_{\kappa}t)S_{2\kappa-2}((1-2s_{\kappa})t)S_{2\kappa-2}(s_{\kappa}t)\, ,
\end{equation}
where $s_{\kappa} = 1/(2-2^{1/(2\kappa - 1)})$. The total number of matrix exponentials $N_{\mathrm{exp}}$ in $S_{2\kappa}$ is
\begin{equation}\label{eq:suzuki_fractal_exponential_count}
    N_{\mathrm{exp}} = 2(J-1)3^{\kappa - 1} + 1\, .
\end{equation}
Thus, the number of matrix exponentials grows exponentially with the order of the product formula.

To address this exponential growth, Yoshida proposed an alternative construction \cite{Yoshida}:
\begin{equation}\label{eq:yoshida_fractal}
    S^{(m)}(t) =
    \left(
    \prod^{m}_{j=1}{S_2(w_{m-j+1}t)}
    \right)
    S_2(w_0 t)
    \left(
    \prod^{m}_{j=1}{S_2(w_{j}t)}
    \right)\, ,
\end{equation}
where $w_j \in \mathbb{R}$ for $j=0,1,\ldots,m$ are parameters. The total number of matrix exponentials $N_{\mathrm{exp}}$ in a formula constructed this way is
\begin{equation}\label{eq:yoshida_exponential_count}
    N_{\mathrm{exp}} = (4m+2)(J-1)+1 .
\end{equation}
The coefficients $w_j$ are determined by recursively expanding Eq.~\eqref{eq:yoshida_fractal} using the Baker–Campbell–Hausdorff (BCH) formula \cite{BCH} and imposing conditions such that all terms below a certain order $k$ vanish. Yoshida constructed product formulae up to the eighth order \cite{Yoshida}, for which $m=7$.

Building on Yoshida’s approach, Morales \textit{et al.} employed numerical optimization to identify new coefficients and constructed both eighth- and tenth-order formulae \cite{Mauro}. Their eighth-order product formula achieves smaller errors than Yoshida’s, and they also introduced a new tenth-order formula.

\subsection{Cost metric based on the number of matrix exponentials}\label{sec:total_cost_analysis}
We apply QPE to the time-evolution operator $U=e^{iHt}$ and implement this unitary approximately using a product formula. As our basic cost metric, we use the total number of matrix exponentials appearing in this product-formula implementation within the QPE circuit, which serves as a surrogate for the total gate cost.
This connection is direct in the Pauli-term decompositions commonly used for molecular Hamiltonians, where the terms $H_j$ in the product formula of Eq.~\eqref{eq:second_order_product_formula} are typically Pauli strings. In this case, each factor $e^{iH_jt}$ appearing in the product formula is a Pauli rotation. A Pauli rotation can be implemented using Clifford gates and a single-qubit $R_Z$ rotation; in a fault-tolerant setting, the non-Clifford rotation often dominates the cost.

To estimate this quantity, we model the eigenenergy error induced by the product-formula approximation. Let $E_{\mathrm{pf}}$ be the eigenenergy of the approximate time-evolution operator constructed from a $p$th-order product formula containing $N_{\mathrm{exp}}$ matrix exponentials, and let $E$ be the corresponding exact eigenenergy of $H$. We define the error-model parameters $\alpha$ and $p$ by
\begin{equation}\label{eq:product_formula_error_model}
E = E_{\mathrm{pf}} + \alpha t^p, \quad \alpha, \, p\in \mathbb{R}\, .
\end{equation}

Under this model, the total number of matrix exponentials required to achieve a target error $\varepsilon_{\mathrm{E}}$, denoted by $F$, is given by
\begin{align}\label{eq:total_gate_count}
  &F \coloneqq \beta N_{\mathrm{exp}}\left(\frac{1+p}{p \varepsilon_{E} }\right)\left(\frac{\alpha ( 1+p)}{\varepsilon _{E}}\right)^{\tfrac{1}{p}}
\end{align}
This expression is obtained by optimizing the allocation of the target-error budget between QPE and the product-formula error.
The derivation of Eq.~\eqref{eq:total_gate_count} is provided in Appendix~\ref{sec:derivation_of_total_gate_count}.

\section{Results}\label{sec:evaluation_results}
\subsection{Methods}\label{sec:evaluation_method}
In this study, we compare and evaluate the total computational cost of various product formulae when calculating the ground state energy of molecular Hamiltonians using phase estimation with a specified target error. In other words, we compare $F$ defined in Eq.~\eqref{eq:total_gate_count} across different product formulae.
In addition, to capture a compilation-aware cost relevant to fault-tolerant implementations such as the STAR architecture, we also estimate the depth of $R_Z$-rotation layers, denoted by $D_{R_Z}$. We define $D_{R_Z}$ as the number of sequential $R_Z$ layers after compiling the grouped evolutions into a form that is diagonal in the computational basis, allowing rotations on disjoint sets of qubits to be performed in parallel.

The formulae considered are the second-order, fourth-order \cite{4thPF}, eighth-order, and tenth-order product formulae. Among the eighth-order cases, we include both Yoshida’s construction and that of Morales \textit{et al.}, and we also include the tenth-order formula constructed by Morales \textit{et al.} For the second-order case we use Eq.~\eqref{eq:second_order_product_formula}, and for the higher-order formulae we adopt the coefficients listed in Table~\ref{tab:product_formula_coefficients} in Appendix~\ref{sec:product_formula_coefficients}.

We choose one-dimensional hydrogen chains $\rm{H}_n$ with $n=2,\ldots,15$ as benchmark systems.
The internuclear distances are fixed at 1.0~\AA, and we use the STO-3G basis set.
For H$_{15}$, we only compare the second-, fourth-, and eighth-order formulae, due to its high computational demand.
Unless otherwise stated, we set the target energy error to $\varepsilon_{\mathrm{E}} = 1.6\times 10^{-4}\,\mathrm{Ha}$, which is one-tenth of chemical accuracy.
Under the Jordan--Wigner transformation, the number of qubits equals the number of spin orbitals; we denote this quantity by $N$ below.

The overall evaluation procedure is as follows:
\begin{enumerate}
  \item Perform restricted Hartree-Fock (HF) calculations using \texttt{PySCF} \cite{pyscf2018,pyscf2020}, and construct the second-quantized Hamiltonian $H$ using the HF basis. We then translate it into Pauli operators using Jordan-Wigner transformation \cite{JW} with \texttt{OpenFermion} \cite{openfermion2017,openfermion2020}.
  We also calculate the true ground state energy $E$ by performing full configuration interaction calculation using \texttt{PySCF}.
  \item \label{itm:eigenvalue_by_perturbation} Evaluate the error $\delta E$ between $E$ and $E_{\mathrm{pf}}$ for multiple values of $t$, and fit it to the form $\alpha t^p$ using the perturbative method described in Appendix~\ref{sec:perturbation_error_comparison}. In the QPE resource estimates below, we fix $p$ to the formal order of each product formula and fit only $\alpha$. For reference only, Appendix~\ref{sec:alpha_p_by_h_chain} reports the values obtained from an unconstrained fit in which both $\alpha$ and $p$ are treated as free fitting parameters; the resulting values remain of comparable magnitude and follow the same qualitative trends.
  \item Substitute the fitted $\alpha$ together with the fixed value of $p$ into Eq.~\eqref{eq:total_gate_count} to calculate $F$, and also evaluate $D_{R_Z}$. We then compare these costs for each product formula as functions of the target error $\varepsilon_{\mathrm{E}}$.
  \item Analyze how each cost metric scales with system size. Specifically, for systems that achieve the target error $\varepsilon_{\mathrm{E}}$, we fit $F$ and $D_{R_Z}$ as functions of the number of qubits $N$ using the models
  \[
  F(N)=a_F N^{b_F}, \qquad D_{R_Z}(N)=a_D N^{b_D},
  \]
  where $a_F$, $b_F$, $a_D$, and $b_D$ are fitting parameters. We then use the fitted models to extrapolate these costs to larger qubit counts.
\end{enumerate}

Note that, at step (1), we group the fermionic Hamiltonian into mutually commuting terms to reduce the error of product formulae. Specifically,
\[
H=\sum_{i} H_i,\quad H_i=\sum_j P_{ij},\quad [P_{ij},P_{ij'}]=0\, ,
\]
where all Pauli terms within the same group $H_i$ commute. Here, $j'$ denotes an index different from $j$ within the group $H_i$, that is, $j' \neq j$.
We then treat each grouped set $H_i$ as a primitive operator and decompose the time evolution according to Eqs.~\eqref{eq:second_order_product_formula} and~\eqref{eq:yoshida_fractal}.
In our implementation, we use the grouping method described by Inoue \textit{et al.} \cite{Inoue24Grouping}.

At step (3), to estimate the depth of $R_Z$-rotation layers, we compile each commuting group $H_i$ into a form that is diagonal in the computational basis.
For the grouped terms that arise in our implementation, the construction of Inoue \textit{et al.} \cite{Inoue24Grouping} yields a common fermionic basis transformation under which all terms in $H_i$ are mapped to polynomials in number operators.
Under the Jordan--Wigner transformation, this basis change is implemented by a Clifford circuit, and the transformed group becomes a sum of terms containing at most two-body $Z$ operators.
We implement the corresponding diagonal evolution using $R_Z$ rotations generated by these $Z$ and $ZZ$ terms, and define $D_{R_Z}$ as the minimum number of sequential $R_Z$ layers required when rotations acting on disjoint sets of qubits are executed in parallel.
We evaluate this layer depth by constructing a conflict graph for the $Z$-type terms and applying a graph-coloring procedure.
In this estimate, we neglect the depth of the Clifford basis-change circuit.
The explicit basis-change construction and the resulting diagonal forms in the computational basis used in our implementation are provided in Appendix~\ref{sec:z_diagonalization_and_rz_layer_depth_estimation}.

\subsection{Results}
Figure~\ref{fig:extrapolation_total_gate_count} shows the total rotation count $F$, while Fig.~\ref{fig:extrapolation_rz_layer_depth} shows the total $R_Z$-layer depth $D_{R_Z}$ under the same target energy error. These figures show that the eighth-order product formula by Morales \textit{et al.} yields the smallest total gate count $F$ for all evaluated H-chains up to $\mathrm{H}_{15}$ and remains optimal in the extrapolation up to 100 qubits. The same formula also attains the smallest depth of $R_Z$-rotation layers over the evaluated systems and throughout the extrapolated qubit range considered.

\begin{figure}[tb]
    \centering
    \includegraphics[width=\linewidth]{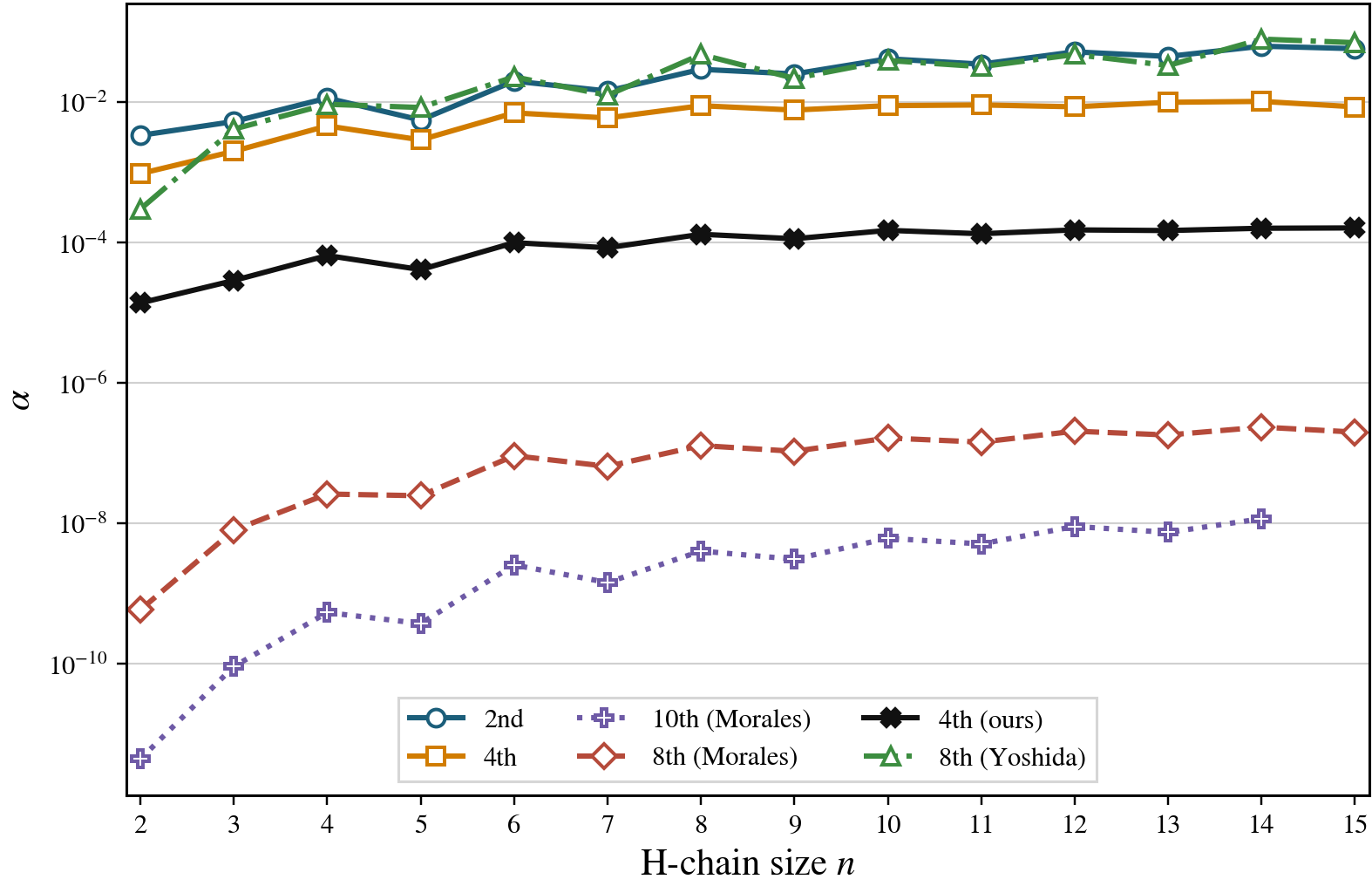}
    \caption{Values of $\alpha$ obtained by fitting with $p$ fixed to the order of each product formula for the H-chain instances considered in this work.}
    \label{fig:alpha_with_fixed_order_by_h_chain}
\end{figure}

The advantage of Morales \textit{et al.}'s formula can be understood more directly by comparing it with Yoshida's formula at the same order. Both are eighth-order formulae within the Yoshida-type construction of Eq.~\eqref{eq:yoshida_fractal}, but Yoshida's formula uses $m=7$, whereas Morales \textit{et al.}'s formula uses one additional pair of second-order blocks, corresponding to $m=8$. Equation~\eqref{eq:yoshida_exponential_count} therefore implies that Morales \textit{et al.}'s formula has a larger per-step decomposition count $N_{\mathrm{exp}}$. However, Fig.~\ref{fig:alpha_with_fixed_order_by_h_chain} shows that it has a much smaller error prefactor $\alpha$. In the total cost formula Eq.~\eqref{eq:total_gate_count}, this reduction in $\alpha$ compensates for the increase in $N_{\mathrm{exp}}$, leading to a smaller total gate count.
The same trade-off also explains why Morales \textit{et al.}'s tenth-order formula does not minimize $F$ in the regime considered here: the additional reduction in error does not offset its larger per-step decomposition count.

\begin{figure}[tb]
    \centering
    \includegraphics[width=\linewidth]{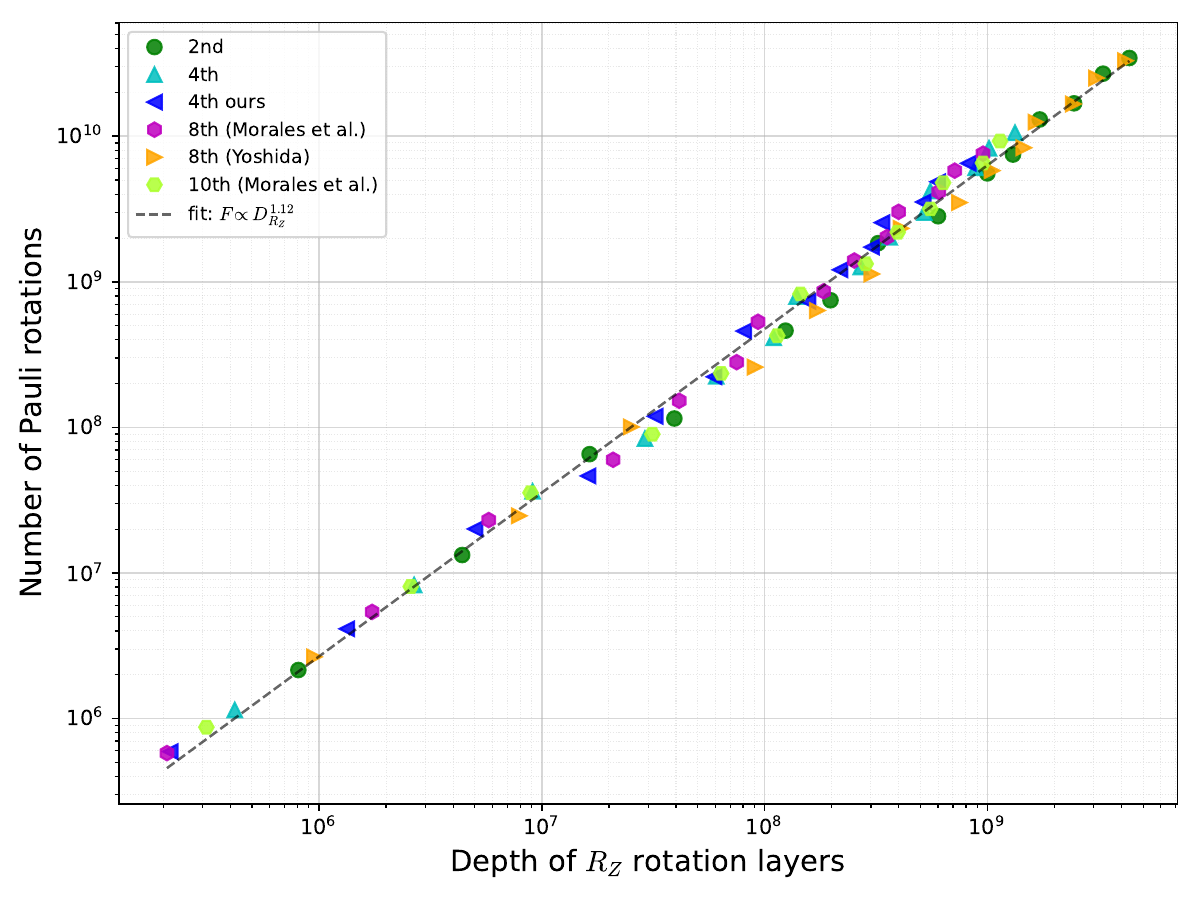}
    \caption{Relationship between the total gate count $F$ and the depth of $R_Z$-rotation layers $D_{R_Z}$ for the H-chain instances considered in this work at the target energy error $\varepsilon_{\mathrm{E}} = 1.6\times 10^{-4}\,\mathrm{Ha}$. The straight line shows the log--log fit to the plotted data.}
    \label{fig:rz_depth_vs_rotations}
\end{figure}

\subsection{Relation between $F$ and $D_{R_Z}$}
\begin{figure*}[tb]
    \centering
    \includegraphics[width=\textwidth]{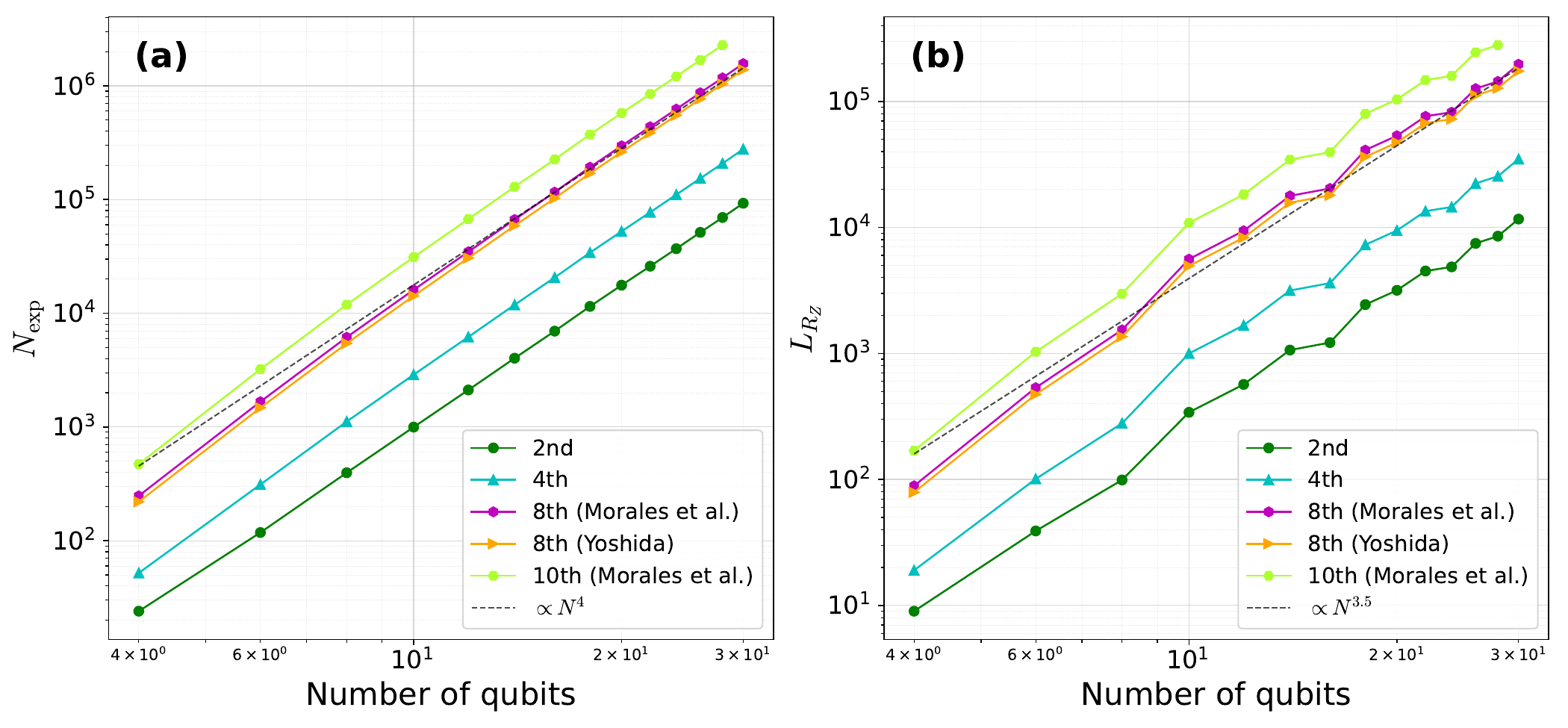}
    \caption{System-size scaling of the per-step quantities for the H-chain instances considered in this work: (a) $N_{\mathrm{exp}}$ and (b) $L_{R_Z}$.}
    \label{fig:nexp_lrz_scaling}
\end{figure*}

Motivated by the preceding benchmark results, in which the same product formula minimizes both $F$ and $D_{R_Z}$, we next examine the relationship between these two cost metrics directly. This allows us to assess whether the total gate count $F$ can serve as a useful proxy for the compilation-aware metric $D_{R_Z}$. Figure~\ref{fig:rz_depth_vs_rotations} shows this relationship for the H-chain instances considered in this work.
On the log--log plot in Fig.~\ref{fig:rz_depth_vs_rotations}, the data lie close to a straight line, indicating an empirical relation close to proportionality between the two metrics. A log--log fit over all H$_2$--H$_{15}$ data points and all product formulae gives
\[
F \propto D_{R_Z}^{1.12},
\]
with Pearson correlation coefficient $r \approx 0.998$. The exponent is close to, but slightly larger than, unity. This can be understood from the corresponding per-step quantities. If we denote by $L_{R_Z}$ the number of $R_Z$ layers in a single product-formula step, then Fig.~\ref{fig:nexp_lrz_scaling} shows the observed system-size scalings $N_{\mathrm{exp}} \sim N^{4}$ and $L_{R_Z} \sim N^{3.5}$. Since $F=rN_{\mathrm{exp}}$ and $D_{R_Z}=rL_{R_Z}$ with the same step number $r$, these imply
\[
F \sim D_{R_Z}^{4/3.5} = D_{R_Z}^{1.14},
\]
in good agreement with the fitted exponent above.

The observed scaling of $L_{R_Z}$ can be interpreted in terms of the per-step costs. Since $D_{R_Z}=rL_{R_Z}$, we have $F/D_{R_Z}=N_{\mathrm{exp}}/L_{R_Z} \sim \sqrt{N}$. In other words, the average number of $R_Z$ rotations packed into one layer scales as $O(\sqrt{N})$, reflecting the effectiveness of parallelization within each commuting group. A naive lower bound from the grouping structure and the graph-coloring formulation in Appendix~\ref{sec:z_diagonalization_and_rz_layer_depth_estimation} gives $L_{R_Z}=\Omega(N^3)$: under the projective-plane-based grouping \cite{Inoue24Grouping}, each group contains on the order of $N^2$ terms, each term expands to $O(1)$ diagonal $Z$- or $ZZ$-rotations, and the resulting conflict graph has maximum degree at least $\Omega(N)$ for each group. Since the number of groups scales as $N^2$, this yields the lower bound $L_{R_Z}=\Omega(N^3)$. The observed scaling $L_{R_Z}\sim N^{3.5}$ therefore suggests an additional $\sqrt{N}$ growth in the chromatic number beyond the maximum-degree bound, which can be interpreted as a structural multiplicity of $O(\sqrt{N})$ overlapping rotations within a group. It will be an interesting direction for future work to determine whether this additional $\sqrt{N}$ overhead is intrinsic to the present grouping and compilation scheme, or whether it can be reduced by improved grouping strategies, diagonalization procedures, or graph-coloring methods.

Overall, these observations indicate that minimizing $D_{R_Z}$ is nearly equivalent to minimizing $F$ in the present benchmark. Under the grouping and compilation scheme adopted here, $F$ can therefore serve as a useful proxy for the compilation-aware metric $D_{R_Z}$. This may be useful in future large-scale surveys of product formulae, where $F$ is substantially cheaper to evaluate than $D_{R_Z}$. At the same time, it will be important to examine in future work whether the same relation persists under different grouping strategies, compilation methods, and hardware assumptions.

\subsection{A new fourth-order product formula}
The benchmark results in Figs.~\ref{fig:extrapolation_total_gate_count} and~\ref{fig:extrapolation_rz_layer_depth} suggest a natural next question. In the regime considered here, the tenth-order formula of Morales \textit{et al.} \cite{Mauro} does not outperform their eighth-order formula. This indicates that increasing the formal order does not automatically reduce the overall cost.

This point can be understood from Eq.~\eqref{eq:total_gate_count}. The total gate count $F$ depends on both the number of exponentials in one product-formula step, $N_{\mathrm{exp}}$, and the error parameters of the formula. In particular, for $\alpha>0$ and $p>0$, reducing $\alpha$ always decreases the total gate count. On the other hand, a direct calculation shows that, if $N_{\mathrm{exp}}$ is held fixed, then $\partial F/\partial p \ge 0$ whenever $\alpha/\varepsilon_E \le 1/(1+p)$. Thus, when the error prefactor is sufficiently small, increasing the formal order $p$ alone does not necessarily reduce the total gate count. Therefore, a higher-order formula is not automatically preferable: if a lower-order construction can retain a sufficiently small error prefactor $\alpha$ while reducing $N_{\mathrm{exp}}$, it can yield a smaller total cost in the practically relevant target-error regime.

The numerical results suggest that the methodology of Morales \textit{et al.} \cite{Mauro} is promising from this perspective. Fig.~\ref{fig:alpha_with_fixed_order_by_h_chain} shows that, among the eighth-order formulae, Morales \textit{et al.}'s construction has much smaller $\alpha$ than Yoshida's across the H-chain instances considered. This observation motivates us to search for a lower-order product formula constructed in the style of Morales \textit{et al.}, with the aim of retaining a small $\alpha$ while reducing $N_{\mathrm{exp}}$.

Therefore, we constructed a fourth-order product formula using Morales \textit{et al.}’s methodology \cite{Mauro} within the Yoshida-type construction of Eq.~\eqref{eq:yoshida_fractal}. Our new formula corresponds to the case $m=2$, with parameters $w_1 = 0.42008729$ and $w_2 = 0.40899193$. As in Morales \textit{et al.}, these parameters were obtained by minimizing the squared residual between the Taylor expansions of $e^{(A+B)t}$ and $S^{(m)}(t,w_1,\ldots,w_m)$ below the target order. The optimized residual is on the order of $10^{-16}$ in our construction, whereas Morales \textit{et al.} reported a threshold below $10^{-22}$ for their eighth- and tenth-order formulae.

Figures~\ref{fig:extrapolation_total_gate_count_fourth_vs_eighth} and~\ref{fig:extrapolation_rz_layer_depth_fourth_vs_eighth} show that our newly constructed fourth-order formula improves upon Morales \textit{et al.}'s eighth-order formula in both cost metrics at the target energy error $\varepsilon_{\mathrm{E}} = 1.6\times 10^{-4}\,\mathrm{Ha}$. Over the evaluated H-chains, it yields both a smaller total gate count $F$ and a smaller depth of $R_Z$-rotation layers, and the extrapolated scaling indicates that this advantage persists up to 100 qubits. In particular, at 100 qubits, the new fourth-order formula reduces both the total number of rotations and the $R_Z$-layer depth by approximately 20\% relative to Morales \textit{et al.}'s eighth-order formula. This improvement is consistent with the behavior of the error coefficient: Fig.~\ref{fig:alpha_with_fixed_order_by_h_chain} shows that the range of $\alpha$ for the new fourth-order formula is about $10^{-5}$ to $10^{-4}$, whereas for the conventional fourth-order formula it is $10^{-4}$ to $10^{-2}$.

\begin{figure}[tb]
    \centering
    \includegraphics[width=\linewidth]{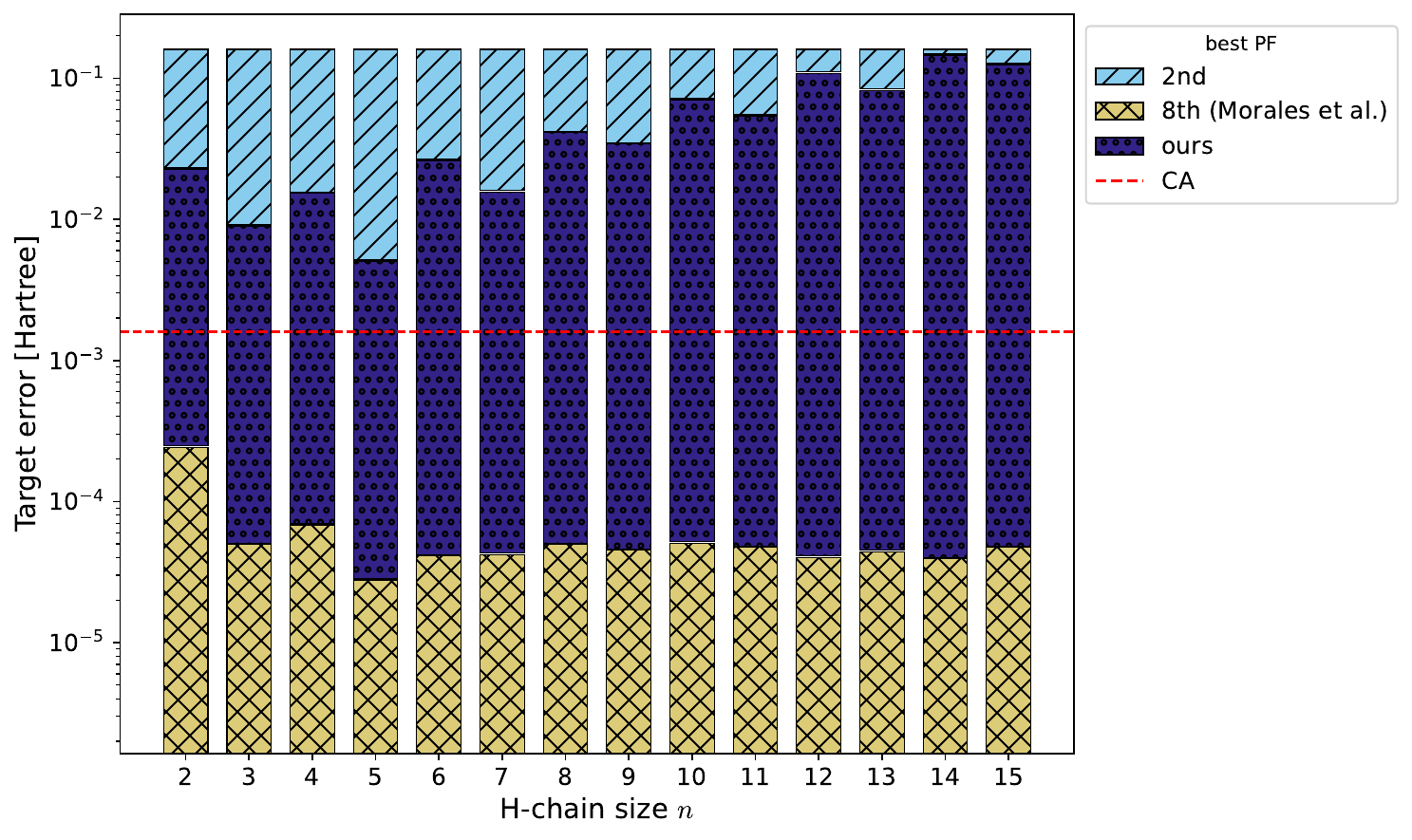}
    \caption{Best (lowest gate count) product formula for each H-chain size $n$ (horizontal axis) and target error (vertical axis, Hartree, log scale). Compared formulae are the second- and fourth-order formulae, the eighth-order formula of Ref.~\cite{Yoshida}, the eighth- and tenth-order formulae of Ref.~\cite{Mauro}, and the fourth-order formula constructed in this work. The red dashed line marks chemical accuracy (CA, $1.6\times10^{-3}\,\mathrm{Ha}$). Around CA, the formula constructed in this work achieves the lowest gate count; for tighter targets the eighth-order formula of Ref.~\cite{Mauro} becomes optimal, and for looser targets the second-order formula is preferred.}
    \label{fig:best_pf_for_error}
\end{figure}

This advantage is most relevant near chemical accuracy. Figure~\ref{fig:best_pf_for_error} shows, for each H-chain, which product formula minimizes the total gate count as a function of the target error. When the new fourth-order formula is included, it yields the smallest total gate count near chemical accuracy for all hydrogen chains from H$_2$ to H$_{15}$. In the lower-error regime beyond chemical accuracy, the eighth-order formula of Ref.~\cite{Mauro} becomes optimal.

\section{Conclusion}
In this study, we compared the computational cost of several product formulae for estimating molecular ground-state energies using quantum phase estimation with a prescribed target accuracy. As benchmark systems, we considered one-dimensional hydrogen chains from $\mathrm{H}_2$ to $\mathrm{H}_{15}$ and evaluated both the total gate count $F$ and the depth of $R_Z$-rotation layers $D_{R_Z}$.

For the H-chain instances considered here, Morales \textit{et al.}'s eighth-order formula gives the smallest $F$ among the previously considered standard formulae, and the same formula also minimizes $D_{R_Z}$. We also found that increasing the formal order does not automatically reduce the overall cost: in the regime considered here, Morales \textit{et al.}'s tenth-order formula does not outperform their eighth-order formula. These results show that the best product formula is determined by the balance between the one-step decomposition count and the error prefactor, not by the formal order alone.

The relation between $F$ and $D_{R_Z}$ is close to proportional in this benchmark, which suggests that, under the grouping and compilation scheme adopted here, $F$ can serve as a useful proxy for the compilation-aware metric $D_{R_Z}$.

Motivated by the observed behavior of the error coefficient $\alpha$, we further constructed a new fourth-order product formula based on the methodology of Morales \textit{et al.} Around chemical accuracy, this new formula gives the lowest total gate count for all evaluated H-chain instances; at the tighter $0.16\,\mathrm{mHa}$ target, it improves upon Morales \textit{et al.}'s eighth-order formula in both $F$ and $D_{R_Z}$ for $\mathrm{H}_3$--$\mathrm{H}_{15}$ and in the extrapolated larger-system regime. In particular, at 100 qubits, the new fourth-order formula reduces both metrics by approximately 20\% relative to Morales \textit{et al.}'s eighth-order formula.

For future work, it will be important to investigate whether the near-equivalence between $F$ and $D_{R_Z}$ persists under different grouping strategies, compilation methods, and hardware assumptions, and whether the additional $\sqrt{N}$ overhead observed in the scaling of $L_{R_Z}$ can be reduced. It will also be useful to extend the analysis to larger systems, alternative QPE algorithms, and Hamiltonian preprocessing techniques such as double factorization and tensor hypercontraction.

\begin{acknowledgments}
This work is supported by JST COI-NEXT Grant No. JPMJPF2014, JST Moonshot R\&D Grant No. JPMJMS256J, and MEXT Quantum Leap Flagship Program (MEXT Q-LEAP) Grant No. JPMXS0120319794. K.M. is supported by JST FOREST Grant No. JPMJFR232Z, JSPS KAKENHI Grant No. 23H03819, 24K16980, and JST CREST Grant Number JPMJCR24I4.
\end{acknowledgments}

\appendix

\section{Numerical Estimation of the QPE Constant $\beta$}\label{sec:beta_estimation}
We estimate the constant $\beta$ for the standard quantum-Fourier-transform-based implementation of quantum phase estimation (QPE) \cite{Kitaev1995PhaseEstimation,AbramsLloyd1999Eigenvalues}. For simplicity, we assume that the input state is an eigenstate of $U$. In standard QFT-based QPE with \(n=\log_2 T\) ancilla qubits, the total number of applications of controlled-$U$ is \(T=2^n\), and the final measurement yields one of the discrete phase values
\[
\theta_j=\frac{2\pi j}{T}, \qquad j=0,\dots,T-1.
\]
Let \(\varphi\) be the corresponding eigenphase, and let \(\hat{\varphi}(T)\) denote the phase estimate obtained at a given value of \(T\). The probability distribution over these discrete phase values is
\begin{equation}
p\!\left(\theta_j;\varphi,T\right)
\;=\;
\frac{1}{T^{2}}\,
\frac{\sin^{2}\!\bigl(\tfrac{T}{2}\,\Delta_j\bigr)}{\sin^{2}\!\bigl(\tfrac{1}{2}\,\Delta_j\bigr)},
\qquad
\sum_{j=0}^{T-1}p\!\left(\theta_j;\varphi,T\right)=1 .\notag
\end{equation}
for \(j=0,\dots,T-1\), where $\Delta_j=\theta_j-\varphi$. We write the resulting estimation error as $\varepsilon(T)$, and we take the most frequently observed value of $\theta_j$ as the estimator $\hat{\varphi}(T)$.

To estimate $\varepsilon(T)$ numerically, we use \(K=9\) sample points
\[
T_i=2^{n_i},\qquad n_i=7,\ldots,15,
\]
together with \(M=100\) measurement samples per QPE run, \(N_{\mathrm{rep}}=10\) repeated QPE runs for each phase and each \(T_i\), and \(N_{\mathrm{trials}}=1000\) independently sampled phases. In the outer loop, we draw phases \(\varphi_t\) independently and uniformly from \([-\pi,\pi]\) for \(t=1,\dots,N_{\mathrm{trials}}\). For each sampled phase \(\varphi_t\) and each \(T=T_i\), we then repeat the following procedure independently for \(r=1,\dots,N_{\mathrm{rep}}\). We draw \(M\) measurement samples from the above QFT-based QPE distribution and take the most frequently observed discrete phase as the phase estimate \(\hat{\varphi}^{(r)}_t(T_i)\) for that run. We then map both \(\varphi_t\) and \(\hat{\varphi}^{(r)}_t(T_i)\) to the interval \([-\pi,\pi]\) and define
\[
\hat{\varepsilon}^{(r)}_t(T_i)
\;=\;
\bigl|\,\hat{\varphi}^{(r)}_t(T_i)-\varphi_t\,\bigr| .
\]
We average these errors over the \(N_{\mathrm{rep}}\) repeated runs to obtain
\[
\varepsilon_t(T_i)=\frac{1}{N_{\mathrm{rep}}}\sum_{r=1}^{N_{\mathrm{rep}}}\hat{\varepsilon}^{(r)}_t(T_i).
\]
Finally, for each \(T_i\), we define \(\varepsilon(T_i)\) as the median of \(\{\varepsilon_t(T_i)\}_{t=1}^{N_{\mathrm{trials}}}\).

We then fit the model
\[
\log \varepsilon(T)=\log \beta-\log T
\]
to the data \(\{(T_i,\varepsilon(T_i))\}_{i=1}^{K}\) by least squares to obtain a single value of \(\beta\). We obtain \(\beta=1.56\). To estimate the uncertainty, we generate 1000 bootstrap resamples by sampling the set of \(N_{\mathrm{trials}}=1000\) phases with replacement, recompute the median-based \(\varepsilon(T_i)\) for each resample, and refit \(\beta\); the resulting standard error is approximately \(0.02\), with a 95\% confidence interval of \([1.53,1.60]\).

We use the median across sampled phases because the error distribution has a heavy tail. When the sampled phase lies near the boundary of the discrete phase grid, especially near \(\varphi\approx\pm\pi\), aliasing at small \(T\) can cause the estimation error to jump to \(O(1)\). As a result, the arithmetic mean is unstable and can be dominated by a small number of outliers; in our bootstrap analysis, the corresponding 95\% confidence interval broadens to roughly \([1.6,3.0]\). The median is much less sensitive to this tail behavior and therefore provides a more stable estimate of the typical scaling, making it a more appropriate representative value for \(\beta\).

\begin{figure}[tb]
    \centering
    \includegraphics[width=\linewidth]{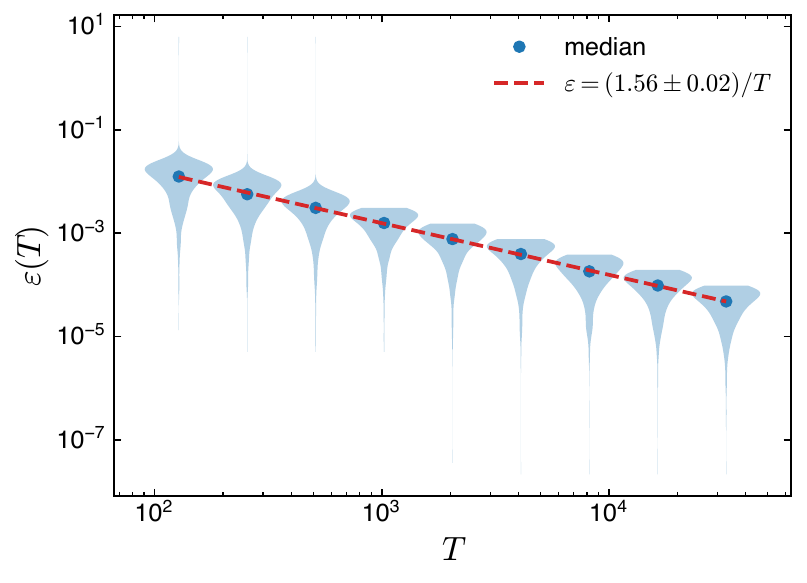}
    \caption{Distribution of the phase-estimation error \(\varepsilon_t(T)\) over \(N_{\mathrm{trials}}=1000\) randomly sampled phases, shown as violin plots for each \(T\). The black markers indicate the median at each \(T\), and the orange dashed line shows the fit to \(\varepsilon=\beta/T\). Here \(N_{\mathrm{rep}}=10\), \(M=100\), and the fitted value is \(\beta=1.56\pm0.02\).}
    \label{fig:qpe_cost_scaling}
\end{figure}

\section{Derivation of the required number of matrix exponentials}\label{sec:derivation_of_total_gate_count}
We consider QPE applied to the target unitary $U=e^{iHt}$. In our implementation, each application of $U$ is approximated by a product-formula circuit, which we denote by $U_{\mathrm{pf}}(t)$ and which contains $N_{\mathrm{exp}}$ matrix exponentials. For the Yoshida-type construction, $N_{\mathrm{exp}}$ is given by Eq.~\eqref{eq:yoshida_exponential_count}. If $M$ denotes the number of applications of this product-formula implementation within the QPE circuit, then the required number of matrix exponentials is
\[
F = M N_{\mathrm{exp}}.
\]
Thus, it is sufficient to determine the minimum value of $M$ needed to achieve a target error $\varepsilon_{\mathrm{E}}$.

To this end, we divide the target-error budget between QPE and the product-formula error. Let $\varepsilon_{\mathrm{QPE}}$ denote the contribution from QPE, and let $\alpha t^p$ denote the contribution from approximating $U$ by $U_{\mathrm{pf}}(t)$, where $t$ is the evolution time and $\alpha, p \in \mathbb{R}$. These two contributions must satisfy
\begin{equation}\label{eq:error_constraint}
    \varepsilon_{\mathrm{QPE}} + \alpha t^p \leq \varepsilon_{\mathrm{E}} \, .
\end{equation}
According to Eq.~\eqref{eq:qpe_cost}, the number of applications required to estimate a phase with precision $\varepsilon$ is $\beta/\varepsilon$. Since QPE estimates the phase $Et$, the relevant phase precision is $\varepsilon_{\mathrm{QPE}} t$, and therefore
\[
M = \frac{\beta}{\varepsilon_{\mathrm{QPE}} t}.
\]
The minimum value of $M$ is then obtained by solving
\begin{equation}
    \mathrm{minimize} \quad \frac{\beta}{\varepsilon_{\mathrm{QPE}} t}
    \quad \mathrm{subject \, to} \quad
    \varepsilon_{\mathrm{QPE}} + \alpha t^p \leq \varepsilon_{\mathrm{E}} \, .
\end{equation}

Using the method of Lagrange multipliers, we find that the optimum is attained on the boundary of the constraint, giving
\begin{align}\label{eq:optimal_evolution_time_and_qpe_error}
    t=\left(\frac{\varepsilon_{\mathrm{E}}}{\alpha (p+1)}\right)^{\tfrac{1}{p}},
    \qquad \varepsilon_{\mathrm{QPE}} = \varepsilon_{\mathrm{E}}\left(\frac{p}{p+1}\right) \, , \\
    \notag \\ \label{eq:minimized_qpe_gate_count}
    \mathrm{minimized} \quad
    \frac{\beta}{\varepsilon_{\mathrm{QPE}} t}
    = \beta \left(\frac{1+p}{p \varepsilon_{\mathrm{E}}}\right)
      \left(\frac{\alpha (1+p)}{\varepsilon_{\mathrm{E}}}\right)^{\tfrac{1}{p}} \, .
\end{align}

Multiplying the minimum value of $M$ by $N_{\mathrm{exp}}$ gives the required number of matrix exponentials, namely Eq.~\eqref{eq:total_gate_count} in the main text.

\section{Eighth- and tenth-order product formulae}\label{sec:product_formula_coefficients}
The parameters of the higher-order product formulae used in this study are listed in Table~\ref{tab:product_formula_coefficients}.

\begin{table*}[tb]
\caption{Parameters of the higher-order product formulae used in this study.}
\begin{ruledtabular}
\begin{tabular}{ccccc}
 & 4th order \cite{4thPF} & Yoshida 8th \cite{Yoshida} & Morales 8th \cite{Mauro} & Morales 10th \cite{Mauro} \\ \hline
$w_1$ & -1.70241438391931 & -1.61582374150097 & 0.29137384767986663096528500968049 & -0.4945013179955571856347147977644 \\
$w_2$ & 1.35120719195965 & -2.44699182370524 & 0.26020394234904150277316667709864 & 0.2904317222970121479878414292093 \\
$w_3$ & - & -0.0071698941970812 & 0.18669648149540687549831902999911 & 0.34781541068705330937913890281003 \\
$w_4$ & - & 2.44002732616735 & -0.40049110428180105319963667975074 & -0.98828132118546184603769781410676 \\
$w_5$ & - & 0.157739928123617 & 0.15982762208609923217390166127256 & 0.98855187532756405235733957305613 \\
$w_6$ & - & 1.82020630970714 & -0.38400573301491401473462588779099 & -0.34622976933123177430694714630668 \\
$w_7$ & - & 1.04242620869991 & 0.56148845266356446893590729572808 & 0.20218952619073117554714280367018 \\
$w_8$ & - & - & 0.12783360986284110837857554950443 & 0.13064273069786247787208895471461 \\
$w_9$ & - & - & - & -0.26441199183146805554735845490359 \\
$w_{10}$ & - & - & - & 0.060999140559210408869096992291531 \\
$w_{11}$ & - & - & - & -0.6855442489606141359108973267028 \\
$w_{12}$ & - & - & - & -0.15843692473786584550599206557006 \\
$w_{13}$ & - & - & - & 0.15414691779958299150286452215575 \\
$w_{14}$ & - & - & - & 0.66715205827214320371061839297055 \\
$w_{15}$ & - & - & - & 0.20411874474696598289603677693511 \\
$w_{16}$ & - & - & - & 0.081207318210272593225087711441684 \\
\end{tabular}
\end{ruledtabular}
\label{tab:product_formula_coefficients}
\end{table*}

\section{Error evaluation by perturbation theory}\label{sec:perturbation_error_comparison}
Let $(E_0,\ket{\psi_0})$ be the ground state energy and ground state of the Hamiltonian $H$. In our setting, these quantities are assumed to be available from a classical quantum-chemistry solver such as \texttt{PySCF}. This makes the perturbative approach useful, because once $(E_0,\ket{\psi_0})$ are known, the product-formula-induced energy error can be estimated without repeatedly diagonalizing the product-formula time-evolution operator itself. Let $U_{\mathrm{pf}}(t)$ be a $p$th-order product-formula approximation of the time-evolution operator.
We define the effective Hamiltonian $H_{\mathrm{eff}}$ by
\begin{align*}
U_{\mathrm{pf}}(t)=e^{-i H_{\mathrm{eff}}t},
\end{align*}
where
\begin{align}
H_{\mathrm{eff}} = H + V
\label{eq:effective_hamiltonian}
\end{align}
for some operator $V$ satisfying $\lVert V \rVert =O\!\big(t^{\,p}\big)$.
Let $E_{\mathrm{pf}}$ denote the ground state energy of $H_{\mathrm{eff}}(t)$, and define
\[
\delta E:=E_{\mathrm{pf}}-E_0.
\]
We define the deviation between the product-formula evolution and the exact evolution as
\begin{align}
\ket{\psi'(t)} &:=U_{\mathrm{pf}}(t)\ket{\psi_0}, \\
\ket{\Delta\psi(t)} &:=\ket{\psi'(t)}-e^{-iE_0 t}\ket{\psi_0}.
\label{eq:definition_of_delta_psi}
\end{align}

By first-order perturbation theory for $H_{\mathrm{eff}}(t)=H+V$, the ground state energy can be approximated as
\begin{align*}
E_{\mathrm{pf}}
= E_0 + \bra{\psi_0}V\ket{\psi_0} + O(\lVert V \rVert ^2).
\end{align*}
Hence
\begin{align}
\delta E = \bra{\psi_0}V\ket{\psi_0} + O(\lVert V \rVert^2),
\label{eq:delta_e_first_order_perturbation}
\end{align}
and substituting $\lVert V \rVert=O(t^{p})$ from Eq.~\eqref{eq:effective_hamiltonian} gives
\begin{align*}
\delta E = \bra{\psi_0}V\ket{\psi_0} + O\!\big(t^{\,2p}\big).
\end{align*}

We are therefore interested in calculating $\bra{\psi_0}V\ket{\psi_0}$.
From Eq.~\eqref{eq:effective_hamiltonian},
\begin{align*}
U_{\mathrm{pf}}(t)&=e^{-i(H+V)t} \\
&= e^{-iHt}\Big(\mathbb{I}-itV+O(t^2\lVert V \rVert)\Big)\\
&=e^{-iHt}\Big(\mathbb{I}-itV+O(t^{p+2})\Big).
\end{align*}
We therefore have,
\begin{align*}
\ket{\psi'(t)}
= e^{-iE_0 t}\Big(\ket{\psi_0}-it\,V\ket{\psi_0}\Big) + O(t^{p+2}),
\end{align*}
and thus,
\begin{align*}
\ket{\Delta\psi(t)}
= -it e^{-iE_0 t} V\ket{\psi_0} + O(t^{p+2}).
\end{align*}
We can take inner product between $\ket{\Delta\psi(t)}$ and $\ket{\psi_0}$ to obtain
\begin{align}
\braket{\psi_0 | \Delta\psi(t)}
&= -ite^{-iE_0 t}\bra{\psi_0}V\ket{\psi_0}
+ O(t^{p+2})\notag \\
&= -ite^{-iE_0 t} \left( \delta E+O(t^{2p}) \right) + O(t^{p+2}) .
\label{eq:ground_state_overlap_with_delta_psi}
\end{align}
Using $e^{-iE_0 t}=\cos(E_0 t)-i\sin(E_0 t)$ and taking the real part of Eq.~\eqref{eq:ground_state_overlap_with_delta_psi}, we obtain
\begin{align*}
\Re \braket{\psi_0 | \Delta\psi(t)}
&= t\,\delta E\,\sin(E_0 t) + O\!\big(t^{2p+1}\big) + O(t^{p+2}).
\end{align*}
Assuming $\sin(E_0 t)\neq 0$, we conclude
\begin{align}
\delta E
&= \frac{\Re\,\braket{\psi_0 |\Delta\psi(t)}}{t\,\sin(E_0 t)}
 + O\!\left(\frac{t^{2p}+t^{p+1}}{\sin(E_0 t)}\right),
\label{eq:delta_e_from_overlap_real_part}
\end{align}
This is the working formula used in our perturbative error estimation.

In the small-\(t\) regime, \(t\sin(E_0t)\simeq E_0t^2\) for \(E_0\neq0\), so the correction term is dominated by \(O(t^p)\). This explains why, for the second-order formula used in the following comparison, the residual difference can show an \(O(t^2)\) behavior. For \(p=2\), this check does not by itself distinguish the exact denominator \(t\sin(E_0t)\) from its small-\(t\) approximation \(E_0t^2\); higher-order formulae would provide a cleaner diagnostic of that distinction.

\begin{figure}[tb]
    \centering
    \includegraphics[width=\linewidth]{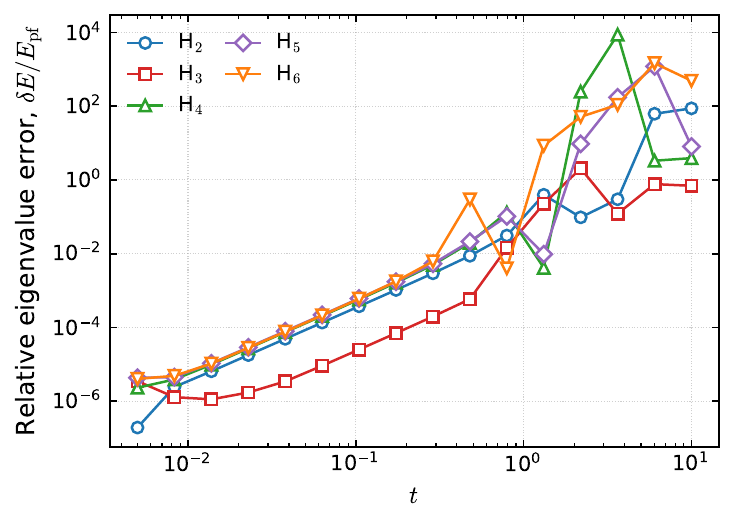}
    \caption{Relative difference between two evaluations of the eigenvalue error $\delta E$ for the $\rm{H_2}$ Hamiltonian with $U_{\mathrm{pf}}(t)$ given by a second-order product formula: direct diagonalization of $U_{\mathrm{pf}}(t)$ and the perturbative estimate based on Eq.~\eqref{eq:delta_e_from_overlap_real_part}.}
    \label{fig:error_direct_vs_perturbation_relative}
\end{figure}

We next compare two ways of evaluating the energy error $\delta E$: one obtained from direct diagonalization of the product-formula evolution operator $U_{\mathrm{pf}}(t)$, and the other obtained from the perturbative estimate based on Eq.~\eqref{eq:delta_e_from_overlap_real_part}. For the $\rm{H_2}$ Hamiltonian with $U_{\mathrm{pf}}(t)$ given by a second-order product formula, Fig.~\ref{fig:error_direct_vs_perturbation_relative} shows the relative difference between these two evaluations. As Fig.~\ref{fig:error_direct_vs_perturbation_relative} indicates, the relative error stays around $10^{-2}$ or smaller for $t \lesssim 10^{-1}$, which is sufficiently small for the present purpose. We therefore use the perturbative estimate of $\delta E$ in the subsequent total-gate-count analysis.

\section{Reduction of grouped Hamiltonian terms to diagonal form in the computational basis and $R_Z$ layer depth estimation}
\label{sec:z_diagonalization_and_rz_layer_depth_estimation}

In this Appendix, we summarize the procedure used to estimate the depth of $R_Z$-rotation layers for each commuting group $H_i$ introduced in Sec.~\ref{sec:evaluation_method}.

We define the fermionic number operator
\begin{equation}
  n_{p\sigma} := a_{p\sigma}^\dagger a_{p\sigma},
\end{equation}
and the Hermitian hopping operator
\begin{equation}
  A_{pq,\sigma} := a_{p\sigma}^\dagger a_{q\sigma} + a_{q\sigma}^\dagger a_{p\sigma}
  \qquad (p\neq q).
\end{equation}
In our implementation, each grouped Hamiltonian term is represented as a linear combination of Hermitian products built from these operators, namely
\begin{align}
  H_i &= \sum_\nu \gamma_{i\nu}\, B_{i\nu},
  \qquad \\
  B_{i\nu} &\in \Bigl\{
  n_{p\sigma},\;
  n_{p\sigma}n_{q\tau},\;
  A_{pq,\sigma}, \notag \\
  &\qquad
  A_{pq,\sigma}n_{r\bar{\sigma}},\;
  A_{pq,\uparrow}A_{rs,\downarrow},\;
  A_{pq,\sigma}A_{rs,\sigma}
  \Bigr\}, \notag
\end{align}
where $\sigma,\tau\in\{\uparrow,\downarrow\}$ and $\bar{\sigma}$ denotes the spin opposite to $\sigma$.

For the commuting groups that appear in our implementation, there exists a fermionic basis transformation that maps the relevant operator products to expressions built from number operators:
\begin{align}
  A_{pq,\sigma}
  &\ \mapsto\ n_{p\sigma}-n_{q\sigma},
  \label{eq:apq_to_z_operator}\\
  A_{pq,\sigma}n_{r\bar{\sigma}}
  &\ \mapsto\ (n_{p\sigma}-n_{q\sigma})n_{r\bar{\sigma}},
  \label{eq:apq_nr_to_z_operator}\\
  A_{pq,\sigma}A_{rs,\tau}
  &\ \mapsto\ (n_{p\sigma}-n_{q\sigma})(n_{r\tau}-n_{s\tau}).
  \label{eq:two_hopping_operators_to_z_operator}
\end{align}
These expressions are diagonal in the occupation-number basis. Under the Jordan--Wigner transformation, this basis transformation is implemented by a Clifford circuit \cite{Inoue24Grouping}, and the number operators are represented by $Z$ operators. The transformed group is therefore implemented by diagonal $Z$-type rotations. For the operator products considered in this work, the expansion contains only constant terms, single-qubit $Z$ terms, and two-qubit $ZZ$ terms. Thus, each group evolution $e^{-itH_i}$ can be implemented using single-qubit $R_Z$ rotations and two-qubit $ZZ$ rotations together with Clifford gates.

We define the depth of $R_Z$-rotation layers as the minimum number of sequential layers required to apply all such rotations when rotations acting on disjoint sets of qubits are executed in parallel.
To estimate this layer depth, we construct a \emph{conflict graph} whose vertices correspond to the individual $Z$-type rotation terms, and we connect two vertices if the corresponding rotations act on at least one common qubit.
The required number of $R_Z$ layers is then estimated by a graph-coloring procedure on this conflict graph.
In this estimate, we neglect the depth contribution from the Clifford basis-change circuit $U_i$ and only count the layers of the diagonal $Z$-type rotations.

\section{$\alpha, \, p$ for each H-chain}\label{sec:alpha_p_by_h_chain}
\begin{figure}[t]
    \centering
    \includegraphics[width=\linewidth]{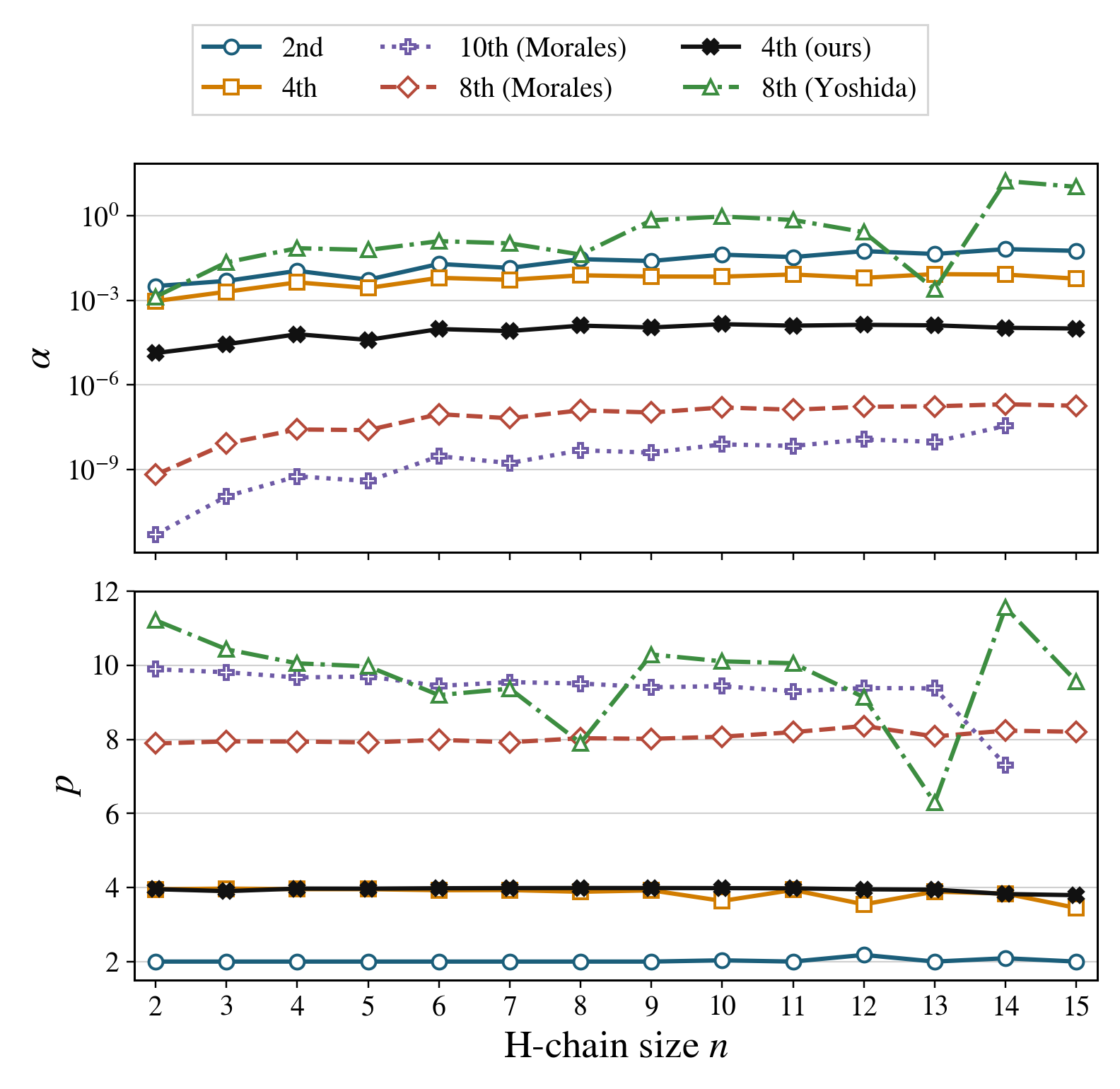}
    \caption{Reference values of the error-model parameters $\alpha$ and $p$ for each H-chain and each product formula, obtained from an unconstrained fit in which both parameters are treated as free fitting parameters. This differs from Fig.~\ref{fig:alpha_with_fixed_order_by_h_chain} in the main text, which is used in the QPE resource estimates and fixes $p$ to the formal order of each product formula while fitting only $\alpha$. Panel (a) shows $\alpha$ on a logarithmic scale, and panel (b) shows $p$.}
    \label{fig:alpha_p_by_h_chain}
\end{figure}

For reference, Fig.~\ref{fig:alpha_p_by_h_chain} shows the fitted values of $\alpha$ and $p$ for each H-chain and each product formula obtained from an unconstrained two-parameter fit. Unlike Fig.~\ref{fig:alpha_with_fixed_order_by_h_chain} in the main text, which is used in the QPE resource estimates and fixes $p$ to the formal order of each product formula, here both $\alpha$ and $p$ are treated as fitting parameters.

\bibliographystyle{apsrev4-2}
\bibliography{ref}

\end{document}